\documentclass[conference]{IEEEtran}
\IEEEoverridecommandlockouts
\pdfoutput=1
\usepackage{cite}
\usepackage{adjustbox}

\usepackage{amsmath,amssymb,amsfonts}
\usepackage{algorithmic}
\usepackage{graphicx}
\usepackage{textcomp}
\usepackage{booktabs}
\usepackage{array}
\usepackage{xcolor}
\usepackage[%
    colorlinks=true,
    pdfborder={0 0 0},
    linkcolor=blue,
    urlcolor=black
]{hyperref}
\usepackage{multirow}

\graphicspath{ {images/} }

\newcolumntype{N}{>{\centering\arraybackslash}m{.1in}}
\newcolumntype{n}{>{\centering\arraybackslash}m{.15in}}
\newcolumntype{G}{>{\centering\arraybackslash}m{0.5in}}

\def\BibTeX{{\rm B\kern-.05em{\sc i\kern-.025em b}\kern-.08em
    T\kern-.1667em\lower.7ex\hbox{E}\kern-.125emX}}
\begin{document}

\title{Pathological Voice Classification Using Mel-Cepstrum Vectors and Support Vector Machine}



\author{\IEEEauthorblockN{Maryam~Pishgar$^{1}$,
Fazle~Karim$^{1}$,~\IEEEmembership{Graduate Student Member,~IEEE},
\\
        Somshubra~Majumdar$^{2}$,
        and Houshang~Darabi$^{1}$,~\IEEEmembership{Senior Member,~IEEE}}
\thanks{$^{1}$Mechanical and Industrial Engineering, University of Illinois at Chicago, Chicago,IL}
\thanks{$^{2}$Computer Science, University of Illinois at Chicago, Chicago, IL}
}

\maketitle

\begin{abstract}
Vocal disorders have affected several patients all over the world. Due to the inherent difficulty of diagnosing vocal disorders without sophisticated equipment and trained personnel, a number of patients remain undiagnosed. To alleviate the monetary cost of diagnosis, there has been a recent growth in the use of data analysis to accurately detect and diagnose individuals for a fraction of the cost. We propose a cheap, efficient and accurate model to diagnose whether a patient suffers from one of three vocal disorders on the FEMH 2018 challenge.

\end{abstract}


\section{Introduction}
The human standard of life can be severely affected by their individual pathological voice condition. This has also financially burdened several patients, organizations, and societies \cite{fang2018detection}. Some of the common impairments to the voice are structural lesions, neoplasms, and neurogenic disorders \cite{fang2018detection}. One of the most frequently utilized tools to diagnose these vocal disorders is through a laryngoscope \cite{denipah2017acute}. Laryngoscopy is an expensive time-consuming process, that requires trained personnel to perform the test \cite{karippacheril2014inexpensive}. In addition, vocal disorders must also be detected and treated at an early diagnostic stage when they are faced with several symptomatic challenges to properly detect, as in the case of larynx cancer \cite{godino2006dimensionality}. Therefore, patients who do not have easy access to advanced technology or those who cannot afford expensive medical treatments are in a derogatory position to receive effective treatments. To alleviate these issues, it has become very popular to utilize non-invasive techniques for detecting vocal disorders through eliminating subjective baises \cite{muhammad2017voice}. In this paper, we propose an algorithm to accurately and quickly classify voice disorder using cheap diagnostic tools. 

In the past 60 years, a significant amount of work has been developed in the field of automated speech pathology, which has assisted physicians to diagnose vocal disorders accurately.  In 1960s, the detection of voice quality was measured by the widespread shimmer, jitter and harmonic-to-the-noise ratio technique (HRN) \cite{hecker1971descriptions}. Subsequently, linear discriminant analysis (LDA) was used to discriminate pathological voices from normal ones \cite{umapathy2005discrimination}. In another study, glottal noise parameter was measured primarily to identify vocal disorder \cite{parsa2000identification}. Another common technique extracted the Mel-frequency cepstral coefficient features (MFCC) and fed into a Gaussian mixture model (GMM) \cite{amara2013voice}. More recently, the F-Ratio and Fisher's discriminant ratio was used for feature selection of acoustic analysis \cite{godino2006dimensionality}. \textit{Little et al}., used recurrence and fractal scaling to quadratically analyze disordered vowels. These samples were then bootstrapped to distinguish normal and disordered voices \cite{little2007exploiting}. In the early 2000s, \textit{Arias-Londoño et  al.} combined the information obtained from MFCC and Modulation spectra (MS) to input into a GMM and support vector machine (SVM) classifier \cite{arias2011combining}. \textit{Arjmandi and Pooyan et al.} focused their study on Short-Time Fourier Transform (STFT) and Continuous Wavelet Transform (CWT) to discriminately analyze voice impairments as they used Linear Discriminant Analysis (LDA), Principle Component Analysis (PCA) and an SVM classifier \cite{arjmandi2012optimum}. In May 2014, \textit{Muhammad and Melhem et al.} employed MPEG-7 features to differentiate pathological voice from normal voice \cite{muhammad2014pathological}. 

Deep learning approaches are becoming more commonly used to categorize voice disorders. One such study utilized convolutional layers with Long Short Term Memory (LSTM) recurrent layers on raw audio signals as an end-to-end model classifier \cite{harar2017voice}. Another deep learning approach used the VGG-16 and CaffeNet models for voice disorder classification \cite{alhussein2018voice}. Successively, \textit{Tsao et al.} compared deep neural networks (DNN), GMM, and SVM models that apply MFCC, MFCC+delta and MFCC(N)+delta features from the raw signals \cite{fang2018detection}.  

In this study, we develop a supervised classification model on a range of given voice data samples (obtained from the 2018 Far Eastern Memorial Hospital Voice Disorder Challenge) to detect the pathological defects and classify them into one of the three commonly identified vocal disorders, Neoplasm, Vocal Palsy, and Phonotrauma. The voice samples included samples of 50 normal voice and 150 voice disorders which were characterized by vocal nodules, polyps, and cysts (mutually related to Phonotrauma), glottis neoplasm, and unilateral vocal paralysis \cite{fang2018detection}. Our model utilizes MFCC and MFCC delta features from the raw signals as an input to an SVM classifier. The hyperparameters of the SVM classifier is tuned via a state-of-the-art algorithm developed by researchers at Google (sequential, halving, and classification algorithm \cite{kumar2018shac}). In this study, we compare our model with a deep learning model (Long Short-Term Memory Fully Convolutional Network \cite{karim2018lstm}) and a classical model (XGBoost) to show how our proposed model can perform similarly to more complicated models on the FEMH dataset \cite{femh}. We propose our model be utilized as a baseline for which physicians can diagnose vocal disorder accurately, and other researchers can compare their results against.

The remainder of the paper is structured as follow:  Section \ref{background} contains the background work on which this work is based. Section \ref{methodology} describes the methodology we use to train our proposed models. In section \ref{results}, we discuss the results obtained on our experiments. Finally, in Section \ref{conclusion}, we conclude and describe potential future work that needs to be performed.

\section{Background Work}
\label{background}
\subsection{Mel-frequency Cepstral Coefficients}

Mel-frequency Cepstrum is a representation of a sound signal, based on the linear cosine transform of a log power spectrum on a nonlinear mel scale of frequency \cite{beigi2011speaker}. The coefficients that collectively comprise the Mel-frequency Cepstrum are called MFCC features. In contrast to the linearly-spaced frequency bands obtained from the cepstrum of a sound signal, in a mel-frequency cepstrum, the frequency bands are uniformly spaced on the mel scale. This frequency warping allows for better representation of sound and voice data.


\subsection{Temporal Derivatives}

In order to extract the dynamic features of speech, auxiliary delta and delta-delta features must be computed as input features for the model, which are calculated as the temporal derivatives of the original MFCC features \cite{krishnan2013sgfiltering}. In order to estimate smooth derivatives $f_{l, n}^{'}$, we often simply compute the local least squares polynomial fit to the data samples, so as to minimize the cost function 

\begin{equation*}
    C_{l,n}^p = \sum_{m=-M}^M \left( \sum_{k=0}^p a_k m^k - f_{l,n+m} \right)^2
\end{equation*}

with respect to the MFCC coefficients $a_k$ where \textit{f} is the spectrum-based feature vector, \textit{L} being the number of coefficients, \textit{N} being the number of frames, \textit{p} is the order of the polynomial and \textit{M} is the number of samples used to fit the polynomial.

We compute the delta features via the local estimate of the derivative of the input data, computed using Savitsky-Golay filtering \cite{krishnan2013sgfiltering} as we utilize the Librosa \cite{mcfee2015librosa} package to preprocess the dataset.  


\section{Methodology}
\label{methodology}
Voice samples were obtained from a voice clinic in a tertiary hospital (Far Eastern Memorial Hospital \textit{(FEMH}). The dataset is comprised of voice samples of 50 individuals who do not exhibit a pathological abnormality in their speech, and voice samples 150 individuals who exhibit one of three different voice disorders, including vocal nodules, polyps, and cysts (collectively referred to \textit{Phonotrauma}); glottis neoplasm and unilateral vocal paralysis. Table \ref{Tab1_Demographics} describes the demographics of the dataset, and Table \ref{Tab2_Diseases} provides the number of samples available for each of the three vocal disorders, stratified by gender.

Each voice sample is a recording of a 3-second sustained vowel sound of the letter "A", under a background noise of amplitude between 40 and 45 decibels, and a microphone-to-mouth distance of approximately 15-20 centimeters. The recordings were obtained using a Shure SM58 microphone and a Shure X2u digital amplifier, with a sampling rate of 44,100 Hz at 16-bit resolution, which was saved in uncompressed waveform-audio format.

\begin{table}[h]
\centering
\caption{\textbf{Demographics of Normal and Pathological Samples}}
\noindent
\begin{tabular}{N N N N N N N N N }\toprule
\multicolumn{1}{N }{\textbf{}} & \multicolumn{2}{c }{\textbf{Count}} &
\multicolumn{2}{c }{\textbf{Mean Age}} &
\multicolumn{2}{c }{\textbf{Age Range}} &
\multicolumn{2}{c }{\textbf{Standard}} \\  

\multicolumn{1}{c }{\textbf{}} & \multicolumn{2}{c }{\textbf{}} &
\multicolumn{2}{c }{\textbf{(Years)}} &
\multicolumn{2}{c }{\textbf{(Years)}} &
\multicolumn{2}{c }{\textbf{Deviation}} \\ 

\cmidrule(lr){2-9}
\multicolumn{1}{c }{\textbf{Gender}} & \textbf{M} & \textbf{F} & \textbf{M} & \textbf{F} & \textbf{M} & \textbf{F} & \textbf{M} & \textbf{F} \\ 

\cmidrule(lr){1-1}
\cmidrule(lr){2-9}

\multicolumn{1}{ c }{\textbf{Normal}} & 20 & 30 & 30.9 & 34.5 & 22-72 & 22-72 & 10.9 & 12. \\ 

\multicolumn{1}{ c }{\textbf{Pathological}} & 78 & 72 & 53.8 & 44.6 & 21-87 & 22-84 & 15.3 & 13.5 \\ 

\bottomrule
\end{tabular}
\label{Tab1_Demographics}
\end{table}

\begin{table}[h]
\centering
\caption{\textbf{Disease Categories of the 150 Pathological Voice Samples}}
\noindent
\begin{tabular}{N N N N}\toprule
\multicolumn{1}{N }{\textbf{}} & \multicolumn{1}{c }{\textbf{Neoplasm}} &
\multicolumn{1}{c }{\textbf{Phonotrauma}} &
\multicolumn{1}{c }{\textbf{Vocal Palsy}}  \\  
\cmidrule(lr){1-4}

\multicolumn{1}{ c }{\textbf{M}} & 32 & 13 & 33 \\ 
\multicolumn{1}{ c }{\textbf{F}} & 8 & 47 & 13 \\ 

\bottomrule
\end{tabular}
\label{Tab2_Diseases}
\end{table}

\subsection{Data Processing}
Each audio waveform is processed to derive MFCC features using the Librosa \cite{mcfee2015librosa} library, using a sampling rate of 22050 Hertz. We compute the temporal derivatives (delta) of these MFCC features, of which we compute the mean and maximum across all samples, and concatenate all three vectors into a single vector of size $\mathbb{R}^{3d}$, where \textit{d} is the number of extracted MFCC coefficients. We select the number of MFCC coefficients (\textit{d}) computed to be 15, which we find via grid search over the space of \textit{\{10, 15, 20, 25, 30, 40, 50, 100\}}.

When assessing the performance of any model, we compute a weighted average of the sensitivity and specificity of the model on the binary task of predicting whether the sample is normal or pathological, and the average recall of the model on the multitask objective of classifying the sample as one of four classes. We assign the weights for the above scores as 40\%, 20\%, and 40\% respectively. 

\begin{figure*}[t]
    \centering
    \includegraphics[width=0.75\linewidth]{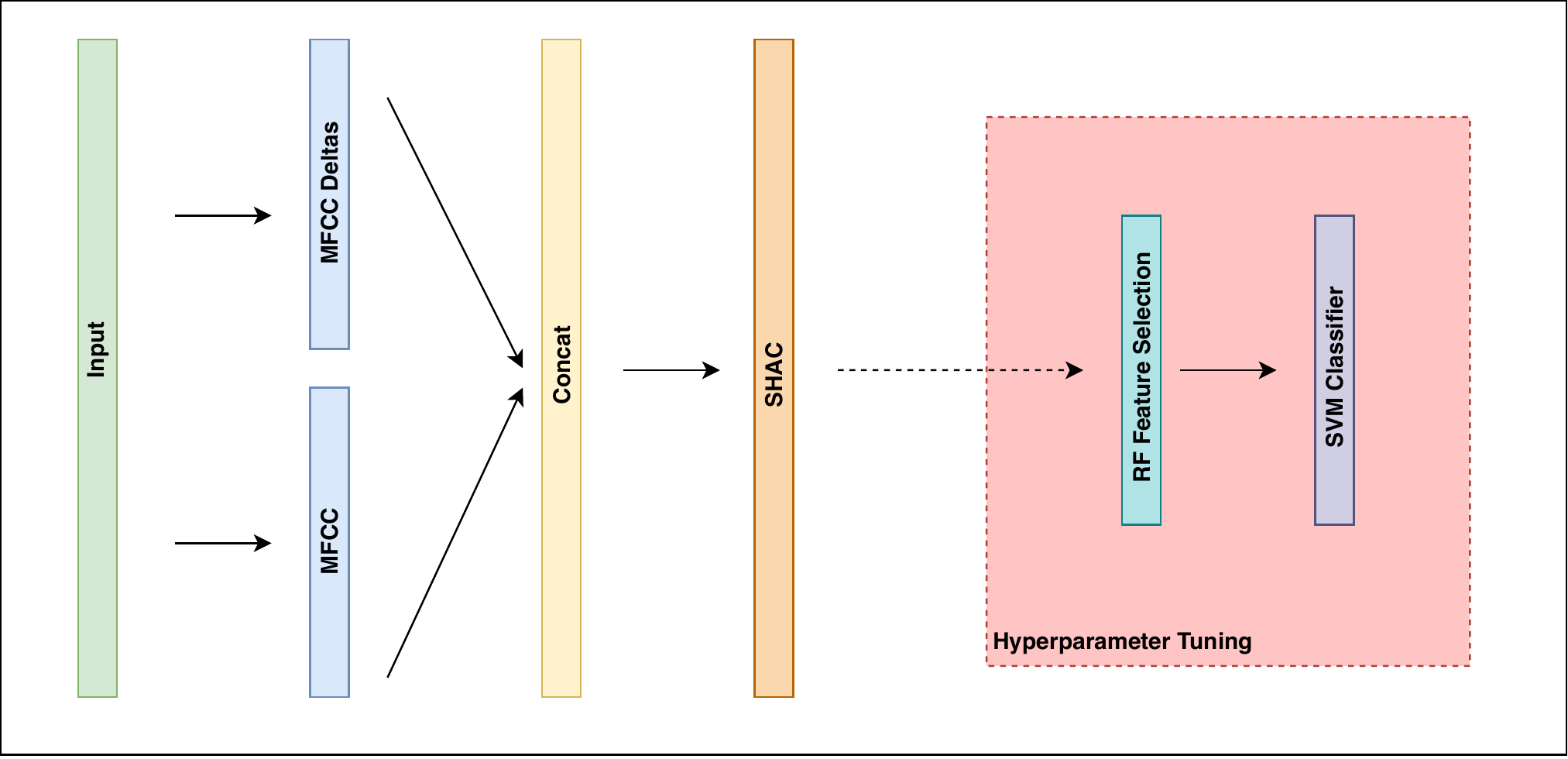}
    \caption{Proposed model pipeline.}
    \label{fig:model_pipeline}
\end{figure*}

\subsection{Hyperparameter Tuning} \label{Hyperparam_tuning}
As the dataset comprises of a mere 200 samples, we perform 5 fold cross validation for every model using the same global seed across all models. We utilize the sequential halving and classification \textit{(SHAC)} algorithm from \textit{Kumar et al.} \cite{kumar2018shac}, as an efficient alternative to exhaustive grid search to sample hyperparameter settings in continuous search spaces. To avoid overfitting the data distribution in the 5 folds used for evaluation of the models, we train and evaluate the \textit{SHAC} algorithm on 5 different folds of the dataset using a different random seed. This preserves the generalization properties of the parameters sampled from the \textit{SHAC} algorithm while also providing robustness to random seed overfitting. For better robustness, we round off all floating point values to the 3rd decimal place, and find that performance is not impacted.

When training the classifiers of the \textit{SHAC} algorithm, we compute a maximum of 10 classifiers, with a batch of 100 hyperparameter samples per classifier and a total budget of 1000 hyperparameter samples evaluated. It is to be noted that once we obtain a sample from the \textit{SHAC} algorithm, we use the same parameters for all 5 folds. Therefore, the sampled parameters are more robust and generalize better, as they must perform well across all 5 folds. We then obtain 100 hyperparameter samples, compute the mean and standard deviation of this batch, and select those candidates which obtain a score greater than the mean plus one standard deviation. We use a publically available implementation of the \textit{SHAC} algorithm \footnote{\href{https://github.com/titu1994/pyshac}{https://github.com/titu1994/pyshac}}.

\subsection{Proposed Model} \label{Proposed_Model}
Our proposed model is comprised of a feature selection stage followed by a multi-class classification stage. We utilize a set of Random Forest models, each trained on the training data of a given fold, to compute the feature importance of the 45 dimension input vector. We then determine a threshold value sampled from the \textit{SHAC} algorithm, which is used to select only those features which are above that threshold. These features are then supplied to a Kernel Support Vector Machine, which utilizes the Gaussian Radial Basis Function as its kernel. We utilize the One-vs-One strategy for multi-class classification, which builds $\frac{N * (N-1)}{2}$ classifiers. Such a strategy is applicable here due to the small amount of data available and relatively fast training, but we observe that One-vs-All strategy is nearly identical in performance. We utilize the excellent Scikit Learn \cite{pedregosa2011scikit} package for all models described in this section. Figure \ref{fig:model_pipeline} details the various stages of the proposed model pipeline.

We construct a search space consisting of the number of trees in the Random Forest \textit{\{10, 20, 50, 100\}}, depth of the tree \textit{\{3, 4, 5, 6, 7, 8, no limit\}}, selection threshold ($U\in(0, 0.5)$), penalty parameter \textit{C} for the SVM ($U\in(0, 25)$), gamma parameter for the RBF kernel ($U\in(-1, 1)$), where we resolve negative values to be \textit{1 / (number of features)}. We then search over this space using the \textit{SHAC} algorithm and sample its best parameters as described in Hyperparameter tuning \ref{Hyperparam_tuning}.

\subsection{Baseline Models}
All baseline models utilize the same input features and 5 fold training as described in the proposed model, to ensure a consistent training methodology. Random Forest based feature selection is performed, but the hyperparameters for the baseline models are searched via \textit{SHAC} to ensure we obtain unbiased hyperparameters. The search space for the Random Forest and the selection threshold remain consistent across all models.

\subsubsection{XGBoost}
We compare against XGBoost \cite{chen2016xgboost}, a powerful Gradient Boosting Tree model which obtains state-of-the-art results on multiple structured and unstructured datasets and is a standard baseline model to compete with.

While XGBoost has a large number of hyperparameters that can be tuned, we find that only three of these parameters significantly impact the final score, and therefore construct a search space only over those three parameters. We search over the number of estimators \textit{\{10, 25, 50, 100, 200\}}, the maximum depth of the tree \textit{\{3, 4, 5, 6, 7, 8\}}, and the learning rate $U\in(0.01, 0.2)$.

\subsubsection{Long Short Term Memory Fully Convolutional Networks}

We also compare this dataset on a hybrid deep neural network, called the Long Short Term Memory Fully Convolutional Network (LSTM-FCN) \cite{karim2018lstm}. It comprises of two branches, one with Convolutional blocks comprised of a Convolutional Layer, followed by Batch Normalization \cite{ioffe2015batch} and then the Relu activation function. Another branch is comprised of a Long Short Term Memory Recurrent layer \cite{hochreiter1997lstm} followed by a dropout \cite{srivastava2014dropout} layer (with a probability of 80\%). 

We utilize the same FCN and LSTM branch structure as provided by \textit{Karim et al.} to be consistent, and only modify the number of LSTM cells in the LSTM branch, which we find using grid search \cite{karim2018lstm}. All aspects of initialization and training methodology are kept consistent with the paper to provide the best results.

\section{Results}
\label{results}
Due to the lack of a distinct test set, we instead discuss the 5 fold cross validation score obtained by each of the models, and how we selected the best model to submit to the FEMH 2018 Challenge. As discussed in \ref{Hyperparam_tuning}, we use the \textit{SHAC} algorithm to randomly sample the best parameters which it determines can obtain strong scores on the overall validation sets of the 5 folds. For each of the models, out of the 100 hyperparameter settings sampled, we found one or more hyperparameter settings which scored above the mean + 1 standard deviation threshold we had set. We then train and evaluate each of these models on those candidates, and rank each of them. We find that we need to train fewer than 3 candidates in all cases as the candidates all score very close to each other. All code employed in this paper is available online. \footnote{\href{https://github.com/houshd/FEMH}{https://github.com/houshd/FEMH}}

The scores of each of these models has been provided in Table \ref{Tab3_Scores}. The final column contains the weighted score of the Sensitivity, Specificity and Recall scores with the weights 40\%, 20\% and 40\% respectively.

\begin{table}[h]
\centering
\caption{\textbf{5-Fold Cross Validation Scores}}
\noindent
\begin{adjustbox}{width = 1 \linewidth}
\begin{tabular}{N N N N n N}\toprule
\multicolumn{1}{c }{\textbf{Model}} &
\multicolumn{1}{c }{\textbf{Sensitivity}} &
\multicolumn{1}{c }{\textbf{Specificity}} &
\multicolumn{1}{c }{\textbf{Recall}} &
\multicolumn{1}{c }{\textbf{Scores}} &
\multicolumn{1}{c }{\textbf{Std. Dev}} 
\\  
\cmidrule(lr){1-6}

\multicolumn{1}{ c }{\textbf{Proposed Model}} & 0.8860 & 0.7823 & 0.5900 & 0.7469 & \textbf{0.0160} \\ 
\multicolumn{1}{ c }{\textbf{XGBoost}} & 0.8747 & 0.7561 & 0.6150 & \textbf{0.7470} & 0.0710 \\ 
\multicolumn{1}{ c }{\textbf{LSTM-FCN}} & 0.8539 & 0.6624 & 0.5550 & 0.6960 & 0.0401 \\ 

\bottomrule
\end{tabular}
\end{adjustbox}
\label{Tab3_Scores}
\end{table}

Upon inspection of the results, we find that the proposed model and XGBoost perform quite similarly. However, the standard deviation of the XGBoost model is much higher than that of the proposed model, albeit the mean of the score is marginally higher. It is for this reason that we state that the proposed model is the overall winner, for scoring marginally higher in sensitivity and specificity, as well as having a lower standard deviation in the weighted score.

We also inspect the reason for the significantly lower performance of the LSTM-FCN. We find that the extremely small amount of data provided to such a large neural network, possessing several hundred thousand parameters allows the model to significantly overfit each of the subsets of the 5 folds during training and consequently receive a much lower generalization score. To see if this was indeed the case, we also train the model with a mere fraction of its original parameters and find that, although the overall score improves, it is not significant enough to be comparable to the proposed model.

\section{Conclusion}
\label{conclusion}
In this study, we present a model that classifies vocal disorders using a range of voice data samples obtained from the 2018 FEMH Voice Disorder Challenge. The proposed model extracts MFCC and MFCC delta features from raw input signals and applies them on a SVM classifier. We utilize the SHAC algorithm to tune the parameters of the classifier for optimal performance. Our results of the proposed model perform similarly and in a case outperformed more complicated models, such as XGBoost and LSTM-FCN.  We recommend using this approach to provide a fast yet simple baseline to diagnose the vocal disorder in clinical practice. We leave the extension of this model to automatically diagnose larynx cancer as future work.

\section*{Acknowledgment}
We would like to thank Far Eastern Memorial Hospital for donating the dataset and the organizers of the 2018 FEMH challenge for providing valuable feedback.

This publication was supported by the Grant or Cooperative Agreement Number, T42OH008672, funded by the Centers for Disease Control and Prevention. Its contents are solely the responsibility of the authors and do not necessarily represent the official views of the Centers for Disease Control and Prevention or the Department of Health and Human Services. 


\bibliographystyle{unsrt} 
\bibliography{biblio} 

\begin{thebibliography}{10}

\bibitem{fang2018detection}
Shih-Hau Fang, Yu~Tsao, Min-Jing Hsiao, Ji-Ying Chen, Ying-Hui Lai, Feng-Chuan
  Lin, and Chi-Te Wang.
\newblock Detection of pathological voice using cepstrum vectors: A deep
  learning approach.
\newblock {\em Journal of Voice}, 2018.

\bibitem{denipah2017acute}
Nizhoni Denipah, Christopher~M Dominguez, Erik~P Kraai, Tania~L Kraai, Paul
  Leos, and Darren Braude.
\newblock Acute management of paradoxical vocal fold motion (vocal cord
  dysfunction).
\newblock {\em Annals of emergency medicine}, 69(1):18--23, 2017.

\bibitem{karippacheril2014inexpensive}
John~George Karippacheril, Goneppanavar Umesh, and Venkateswaran Ramkumar.
\newblock Inexpensive video-laryngoscopy guided intubation using a personal
  computer: initial experience of a novel technique.
\newblock {\em Journal of clinical monitoring and computing}, 28(3):261--264,
  2014.

\bibitem{godino2006dimensionality}
Juan~Ignacio Godino-Llorente, Pedro Gomez-Vilda, and Manuel Blanco-Velasco.
\newblock Dimensionality reduction of a pathological voice quality assessment
  system based on gaussian mixture models and short-term cepstral parameters.
\newblock {\em IEEE transactions on biomedical engineering}, 53(10):1943--1953,
  2006.

\bibitem{muhammad2017voice}
Ghulam Muhammad, Mansour Alsulaiman, Zulfiqar Ali, Tamer~A Mesallam, Mohamed
  Farahat, Khalid~H Malki, Ahmed Al-nasheri, and Mohamed~A Bencherif.
\newblock Voice pathology detection using interlaced derivative pattern on
  glottal source excitation.
\newblock {\em Biomedical Signal Processing and Control}, 31:156--164, 2017.

\bibitem{hecker1971descriptions}
Michael~HL Hecker and E~James Kreul.
\newblock Descriptions of the speech of patients with cancer of the vocal
  folds. part i: Measures of fundamental frequency.
\newblock {\em The Journal of the Acoustical Society of America},
  49(4B):1275--1282, 1971.

\bibitem{umapathy2005discrimination}
Karthikeyan Umapathy, Sridhar Krishnan, Vijay Parsa, and Donald~G Jamieson.
\newblock Discrimination of pathological voices using a time-frequency
  approach.
\newblock {\em IEEE Transactions on Biomedical Engineering}, 52(3):421--430,
  2005.

\bibitem{parsa2000identification}
Vijay Parsa and Donald~G Jamieson.
\newblock Identification of pathological voices using glottal noise measures.
\newblock {\em Journal of speech, language, and hearing research},
  43(2):469--485, 2000.

\bibitem{amara2013voice}
Fethi Amara and Mohamed Fezari.
\newblock Voice pathologies classification using gmm and svm classifiers.
\newblock {\em Recent Advances in Biology, Medical Physics, Medical Chemistry,
  Biochemistry and Biomedical Engineering}, page~65, 2013.

\bibitem{little2007exploiting}
Max~A Little, Patrick~E McSharry, Stephen~J Roberts, Declan~AE Costello, and
  Irene~M Moroz.
\newblock Exploiting nonlinear recurrence and fractal scaling properties for
  voice disorder detection.
\newblock {\em Biomedical engineering online}, 6(1):23, 2007.

\bibitem{arias2011combining}
Juli{\'a}n~David Arias-Londo{\~n}o, Juan~I Godino-Llorente, Maria Markaki, and
  Yannis Stylianou.
\newblock On combining information from modulation spectra and mel-frequency
  cepstral coefficients for automatic detection of pathological voices.
\newblock {\em Logopedics Phoniatrics Vocology}, 36(2):60--69, 2011.

\bibitem{arjmandi2012optimum}
Meisam~Khalil Arjmandi and Mohammad Pooyan.
\newblock An optimum algorithm in pathological voice quality assessment using
  wavelet-packet-based features, linear discriminant analysis and support
  vector machine.
\newblock {\em Biomedical Signal Processing and Control}, 7(1):3--19, 2012.

\bibitem{muhammad2014pathological}
Ghulam Muhammad and Moutasem Melhem.
\newblock Pathological voice detection and binary classification using mpeg-7
  audio features.
\newblock {\em Biomedical Signal Processing and Control}, 11:1--9, 2014.

\bibitem{harar2017voice}
Pavol Harar, Jesus~B Alonso-Hernandezy, Jiri Mekyska, Zoltan Galaz, Radim
  Burget, and Zdenek Smekal.
\newblock Voice pathology detection using deep learning: a preliminary study.
\newblock In {\em Bioinspired Intelligence (IWOBI), 2017 International
  Conference and Workshop on}, pages 1--4. IEEE, 2017.

\bibitem{alhussein2018voice}
Musaed Alhussein and Ghulam Muhammad.
\newblock Voice pathology detection using deep learning on mobile healthcare
  framework.
\newblock {\em IEEE Access}, 6:41034--41041, 2018.

\bibitem{kumar2018shac}
Manoj Kumar, George~E Dahl, Vijay Vasudevan, and Mohammad Norouzi.
\newblock Parallel architecture and hyperparameter search via successive
  halving and classification.
\newblock {\em arXiv preprint arXiv:1805.10255}, 2018.

\bibitem{karim2018lstm}
Fazle Karim, Somshubra Majumdar, Houshang Darabi, and Shun Chen.
\newblock Lstm fully convolutional networks for time series classification.
\newblock {\em IEEE Access}, 6:1662--1669, 2018.

\bibitem{femh}
Chi-Te Wang, Feng-Chuan Lin, Yu~Tsao, and Shih-Hau Fang.
\newblock {2018 FEMH Voice Data Challenge}, May 2018.
\newblock \url{https://femh-challenge2018.weebly.com/}.

\bibitem{beigi2011speaker}
Homayoon Beigi.
\newblock {\em Speaker recognition}.
\newblock Springer, 2011.

\bibitem{krishnan2013sgfiltering}
S.~R. Krishnan, M.~Magimai.-Doss, and C.~S. Seelamantula.
\newblock A savitzky-golay filtering perspective of dynamic feature
  computation.
\newblock {\em IEEE Signal Processing Letters}, 20(3):281--284, March 2013.

\bibitem{mcfee2015librosa}
Brian McFee, Colin Raffel, Dawen Liang, Daniel~PW Ellis, Matt McVicar, Eric
  Battenberg, and Oriol Nieto.
\newblock librosa: Audio and music signal analysis in python.
\newblock In {\em Proceedings of the 14th python in science conference}, pages
  18--25, 2015.

\bibitem{pedregosa2011scikit}
Fabian Pedregosa, Ga{\"e}l Varoquaux, Alexandre Gramfort, Vincent Michel,
  Bertrand Thirion, Olivier Grisel, Mathieu Blondel, Peter Prettenhofer, Ron
  Weiss, Vincent Dubourg, et~al.
\newblock Scikit-learn: Machine learning in python.
\newblock {\em Journal of machine learning research}, 12(Oct):2825--2830, 2011.

\bibitem{chen2016xgboost}
Tianqi Chen and Carlos Guestrin.
\newblock Xgboost: A scalable tree boosting system.
\newblock In {\em Proceedings of the 22nd acm sigkdd international conference
  on knowledge discovery and data mining}, pages 785--794. ACM, 2016.

\bibitem{ioffe2015batch}
Sergey Ioffe and Christian Szegedy.
\newblock Batch normalization: Accelerating deep network training by reducing
  internal covariate shift.
\newblock {\em arXiv preprint arXiv:1502.03167}, 2015.

\bibitem{hochreiter1997lstm}
Sepp Hochreiter and J{\"u}rgen Schmidhuber.
\newblock Long short-term memory.
\newblock {\em Neural computation}, 9(8):1735--1780, 1997.

\bibitem{srivastava2014dropout}
Nitish Srivastava, Geoffrey Hinton, Alex Krizhevsky, Ilya Sutskever, and Ruslan
  Salakhutdinov.
\newblock Dropout: a simple way to prevent neural networks from overfitting.
\newblock {\em The Journal of Machine Learning Research}, 15(1):1929--1958,
  2014.

\end{thebibliography}

\end{document}